\documentclass[a4paper]{jpconf}
\usepackage{graphicx}
\usepackage{citesort}
\usepackage{color}

\usepackage{epstopdf}

\begin{document}
\title{Open issues in  probing interiors of solar-like oscillating  main sequence stars\\
1. From the Sun to  nearly suns}

\author{MJ Goupil${^1}$,  Y. Lebreton${^1}$, J.P. Marques${^1}$, R. Samadi${^1}$, F. Baudin$^2$}

\address{$^1$ Observatoire de Paris,  UMR 8109, Paris, France}
\address{$^2$ Universit\'e  Paris -Sud, Orsay, France}

\ead{mariejo.goupil]@obspm.fr}

\begin{abstract}
We review some major, open issues in the current modelling 
of low and intermediate mass, main sequence stars based on seismological studies. 
In the present paper, the solar case  is discussed together with current  problems
that are  common to the Sun and  stars with a 
structure similar to that of the  Sun. Several additional issues specific to main sequence 
stars  other than the Sun are reviewed and 
 illustrated with a few stars observed with CoRoT in a companion paper.
\end{abstract}

\section{Introduction}
After   more than two  decades  of helioseismology, 
almost four  years  of asteroseismology with CoRoT \cite{baglin06} and almost two years  
 of  intensive asteroseismology  with Kepler \cite{borucki07}, we review some major, current open
issues about the internal structure of the Sun and solar-like oscillating  stars.
 We   discuss here  the solar case, this also applies to 
 oscillating  stars that have a similar internal structure. 
For sake  of  brevity, we decided  to include  only unpublished figures and to 
cite published figures in the text. Several recent reviews exist on the topic,
 for instance \cite{jcd09}, \cite{jcdhoudek10}, \cite{basuantia08}. 

\section{The Sun }
As it is well known, the Sun is a particular case. 
It is the closest star, and as a result we know with a high precision 
the luminosity, mass (through the product  $G M_\odot$), radius, age 
and individual surface abundances of chemical elements\footnote{although some of 
these latter are still debated, see below}.
Furthermore, a wealth of very accurate seismic constraints are available 
and have been successfully used. Inversion of a large set of  mode 
frequencies has provided crucial information on the structure of the Sun,
  see for instance \cite{basuantia08}; \cite{jcd02}.
  Accordingly, the following constraints must  all be satisfied by any calibrated solar model: 
  radius at the base of the convective envelope $r_{bzc}$, surface helium abundance $Y_s$, 
  sound speed profile $c(r)$, internal rotation profile and location of ionization regions. 
  To some extent, these constraints  are found  to be independent of the reference model  
  \cite{basu2000}. The current  major challenges and open issues  in the solar case then are:
\begin{itemize}
\item what are the values of the surface abundances, more specifically the oxygen abundance?
\item what is the origin of the  discrepancy between the seismic  sound speed and
that given by models below the convection zone?
\item what are the dominant physical mechanisms responsible for 
a uniform rotation in the radiative region and a differential rotation 
in the convective zone? It is worth  noting that this is the opposite in  current calibrated 1D solar models: the
convection zone is assumed to rotate uniformly and  the rotation profile in the  radiative zone is found to vary
with radius if waves and/or magnetic field are not taken into account (see below) !
\item how can we model properly near surface layers  and
 the convection-pulsation interaction?
\item how to succeed in probing the core?
\item how to model oscillation mode line widths and amplitudes?
\end{itemize}

Some of these uncertainties about the Sun 
have consequences on the modelling of stars other than the Sun. On the other
hand,  some problems that are encountered with seismological studies of 
stars other than the Sun can be studied first with  our well
known Sun.   These two points of view are  discussed below.

\subsection{Initial abundances: the solar mixture }

In the late nineties, with the GN93 solar abundances \cite{grevesse93}, 
the seismic sun and calibrated solar models were in agreement by 1 to 5\%  for the sound speed profile and location of 
the base of the convective zone  \cite{basu08}. However between 1993 and 2010 several revisions of the photospheric 
solar mixture were performed. Noteworthy, 3D model atmospheres including  NLTE effects  and  
improved atomic data have led to a substantial 
 decrease of the C, N, O, Ne, Ar abundances and in turn to an 
 important decrease of the solar metallicity $(Z/X)$ (see Table \ref{mixture}).

\begin{table}
\caption{Value of the solar photospheric metallicity from 1993 to 2010 from Table 4 in Asplund \etal\ 
\cite{asplund09} 
complemented by Caffau \etal\ \cite{caffau08} }
\begin{tabular}{lllllll}
\hline
      &GN93 & GS98 & AGS05 & AGS09 & Lod09 &Caff10\\
$Z/X$ &0.0244&0.0231&0.0165&0.0181&0.0191&0.0209\\
\hline
\end{tabular}
\label{mixture}
\end{table}

\begin{figure}[t]
\centering
\resizebox*{0.4\hsize}{!}{\includegraphics*{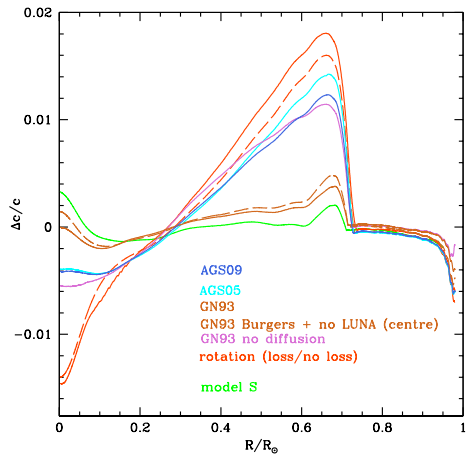}}
\caption{\small  Differences in the sound speed profiles for the Sun between 
Basu's seismic model and calibrated solar models using various abundance
mixtures and different assumptions for the physical description. 
All models have been calculated with the Cesam2k code except  model S  \cite{jcd96} which is
 based on the old solar mixture.}
\label{cs}       
\end{figure}

Today, an internal consistency  of the abundance determination from different ionization levels 
of a given element  seems  to have been obtained and a consensus between independent determinations 
seems to be  reached  (i.e. this shows the unvaluable benefit of independent approaches).
 However the revised modern solar chemical composition \cite{asplund09} leads to a strong disagreement between the 
 sound speed of a calibrated solar  model and that of the seismic Sun. This is also true for  the 
 depth of the convection zone and the envelope helium abundance (see \cite{basuantia08}, 
 for a review). Fig.\ref{cs} shows the difference  of the sound speed profile between
   Basu's seismic model  \cite{basu97}  and several models computed with Cesam2k \cite{morel08} 
   using  some of the  abundance mixtures listed in Tab.\ref{mixture}. This figure shows 
   that the discrepancy at the base of the convection zone is strongly increased when 
   the revised abundance mixture AGS05 is used. This is mainly due to the decrease in 
   oxygen and neon and in $(Z/X)_s$ which consequence is a decrease of the radiative opacities.  
With the newly derived abundance mixture AGS09 which has achieved  some consensus, 
the discrepancy slightly decreases but remains significant.

\subsubsection{Possible origins of the discrepancy}

First there could be some errors in the opacity derivation. As a check, the comparison between OPAL and OP opacities for a 
calibrated solar model shows that OP opacities give a (slightly) better fit than OPAL.
 However a change in opacity of about 30\% at the base of the convection zone, at a temperature of $2~ 10^6$ K 
 would be necessary to compensate for the effects of the change in mixture
  \cite{bahcall05, jcd09} and 
 according to \cite{Badnell05} there appears to be no way to change the OP opacity  by such an amount  
 (see  also \cite{basu10d} and \cite{turck10d}).
Other possible causes have been discussed but none of the related improvements 
can reconcile simultaneously all the seismic constraints listed 
above \cite{guzik06},\cite{guzik08}, \cite{basu07}.

In order to reconcile the seismic and  theoretical sound speed,  one 
needs higher opacities or higher helium below the upper convective zone (UZC) i.e. a higher He gradient. 
Any mixing below the UZC which smooths the gradient therefore would go  in the wrong direction. 
In that case, one would rather 
need an advection process.  It is not clear whether internal waves and/or which hydrodynamical instability  could  act in the sense of increasing helium below the UZC.
Other further developements  are underway such as   taking magnetic effects into account 
 in the derivation of    solar abundance corrections.  An investigation based on  3D magnetoconvection simulations 
     shows that corrections to the solar abundance can be
  significant \cite{fabbian10}.

 Abundances of other stars are determined by reference to the Sun, 
hence all stars are affected by  errors or by inaccuracies in the solar mixture. Can other stars be discriminating in that issue?
Note that one must wonder about the impact of the inconsistency which arises when modeling other 
stars using solar mixtures inferred from 3D model atmospheres if the stellar [Fe/H] itself has not been
 determined from a dedicated 3D model atmosphere.

\subsection{ Nuclear reaction rates}

In stars, (most commonly charged induced) nuclear reactions occur at low energies (10- 300 keV) 
in the Gamow peak -which corresponds to the maximum probability of the reaction. 
The cross sections that are 
governed by  Coulomb barriers and resonances show a strong and complex dependance on energy and globally 
decrease steeply towards low energy. 

Cross sections are usually written in the form 
\begin{equation}
\sigma(E)= {S(E)\over E}   ~ e^{-2\pi \eta}
\label{nuc}
\end{equation}
where $S(E)$  -the astrophysical factor- contains everything that concerns the nuclei and nuclear physics
and varies slowly with $E$. The exponential term is related to the Coulomb barrier and 
relative velocity of the nuclei. Experiments which provide measurements of the cross sections 
in the laboratory generally
occur at higher energy than the Gamow peak. Extrapolation of $S(E)$ to lower  $E$ 
is then necessary. This is difficult to achieve and cannot take into  account  possible 
unknown resonances occuring at low $E$.
When the $S$-factor cannot be measured,  the reaction rates are obtained from pure theory.
Recent significant progress in laboratory and theory (hence better determination of 
 $S$-factors down to the Gamow peak) has been achieved as discussed in the comprehensive
  reviews by \cite{weiss08}, \cite{adelberger10}, see also \cite{lebreton10b}.

\subsubsection{Hydrogen burning reaction rates: }

Uncertainties still exist for the pp chain and CNO cycle reaction cross sections in the Sun. They are due to the difficulty 
to estimate the $S$-factor at low energy and to determine the exact role of the electron screening (both in the laboratory 
and in the star). For most of the concerned reactions, $S$-factors are determined  from extrapolation of experimental data 
to low $E$ but now, for some key reactions, energies corresponding to the Gamow peak are accessible through measurements 
provided by the LUNA experiment at Gran Sasso \cite{costantini09}. This is the case of the $^3He(^3He, 2p) ^4He$ reaction (see Fig. 1 in \cite{costantini09}) for which the $S(E)$-value has not been significantly modified but the error bars are largely decreased.
Furthermore the cross section of the $N^{14}(p, \gamma) O^{15}$ reaction -the leading reaction in the CNO cycle- has now been measured
 down to energies relevant for $1 M_\odot$ stars on the red giant branch (see Fig. 19 in \cite{costantini09}) and a resonance is observed.
  This represents a significant advance  although an extrapolation is still needed towards solar conditions. 
  Noteworthy, the revision is important with a decrease of $S(E)$ by 50\%. This has crucial consequences for solar and stellar structure. In the solar core, the CNO cycle efficiency 
is reduced from $1.6$ to $0.8\%$ of the total energy. For main sequence stars slightly more massive than the Sun the occurrence of
 a convective core depends on the energy production. With the new $N^{14}(p, \gamma) O^{15}$ rate the convective core appears at 
 higher stellar mass and is less massive for a given mass (see for instance the case of a 1.2 $M_\odot$ low metallicity star 
 in Fig. 14 by \cite{lebreton10a}).
Finally cross sections that are obtained from pure theory can be constrained by helioseismology 
\cite{DeglInnocenti98},\cite{weiss01},\cite{weiss08}.

\subsubsection{Electron Screening}

Electron screening is  based on the Salpeter's formula \cite{Salpeter54}
with the underlying physical picture: the cloud of electrons decrease the repulsive Coulomb
 effect between  interacting nuclei
with the result of a decrease of the Coulomb potential
 and an enhancement of the reaction rate. This is a static description. 
 It is currently 
 not clear whether dynamic effects of the interacting ions 
can   significantly change the impact of the  screening 
in nuclear interactions for stars and  wherther it must therefore
be taken into account. 
 The energy that initially fast moving interacting ions have when they get 
  close enough to interact 
 can be lower than the mean value of the medium; accordingly 
 their reaction rate is reduced compared to  Salpeter's prescription. This effect is difficult to  quantify 
and    relies  on results from numerical simulations \cite{shaviv2000}, \cite{shaviv10},  
\cite{mussack10}, \cite{mao09} and the issue is not settled yet. 
\cite{weiss01} looked at the  impact of  changing the electron
screening compared with the classical Salpeter's formula on the solar sound speed
profile. They found that the solar seismic constraints  do not allow variations larger than 1\% when using
GS98.  \cite{jcd09} computed a solar model equivalent to
the S model but  assuming  the extreme case of  no screening at all. 
Their Fig.12 shows  the difference
 in the sound speed profile between the  
seismic sun \cite{basu97} and   model S but switching off  e- screening. 
  A decrease of the reaction rate by swithcing off the e-screening clearly increases the discrepancy between
  observation and model. This is in agreement with previous results of   \cite{weiss08}  who  showed that the large sound speed discrepancy 
with the use of AGS05 mixture is reduced to the previous level  obtained 
with GS98 when the pp-reaction and the screening factor are
 increased up to 15\% (Fig.3 in \cite{weiss08})
In that case,  the surface
  helium and the depth of the convective envelope 
  are in agreement with seismic
  determinations  but the sound speed profile in the core significantly deviates 
  from the seismic solar sound speed.

\subsection{ Transport of chemicals and angular momentum}

{\it Rotationally induced transport}
The physical origin of the  uniform rotation profile in the radiative zone of the Sun 
unravelled by helioseismology   
is still debated. 
Rotationally induced transport of angular momentum
resulting from a competition between 
shear-induced turbulence and meridional circulation driven by surface angular
momentum  losses 
is not able to  make the rotation uniform in the radiative region of the Sun. 
 For details, the reader is refered to \cite{maeder09a}.
The impact of rotationally induced transport  on the solar sound speed  profile 
 has been investigated by  \cite{palacios06}; \cite{yang06}.  \cite{turck10}   have  computed 
 a  solar model including rotationally induced transport 
with two different assumptions about the initial velocity (slow or `fast' sun).
 Their conclusion is that for an initially slow rotation, the microscopic diffusion dominates
whereas for an initially rapid enough rotation,  meridional  circulation dominates over turbulent shear.
However in both cases, the discrepancy below the UCZ  increases 
  (see Fig.9 of \cite{turck10}). This is also illustrated in Fig.\ref{cs} which 
  shows the differences in sound speed profile  for calibrated solar
models computed with
AGS05  mixture computed  assuming either no rotationally induced transport,
or  rotationally induced transport with  no surface angular momentum  loss or 
assuming rotationally induced transport with  surface angular momentum  loss.
 Adding rotationally induced transport makes the
discrepancy with the observed sound speed worse. 
Indeed this process smoothes the helium  gradient below the UZC whereas
 the observations seem to require an increase of
helium in the radiative zone below the UCZ. 
    Several  prescriptions have been derived 
   for the   turbulent transport coefficients involved in rotationally induced transport but  
    validitation  of these prescriptions remain to be done  \cite{Talon04a}.\\
 {\it Internal wave induced transport}  in radiative zones must exist in stars 
as these waves are generated at the interface between convective and
radiative regions. They have been shown to transport angular momentum efficiently enough to 
make  the rotation of the Sun  in  its radiative part rigid \cite{talon05}.  One the main uncertainties 
is  related to the absence of a viable quantitative description of the generation of waves. \\
{\it  Magnetic induced transport} Instabilities driven by the interaction between rotation and 
   a magnetic field     could be 
   responsible for  the transport of angular momentum and  the rigid  solar rotation \cite{maeder09b}. 
 This was shown by \cite{yang06} in the solar case.

Whether all 3 processes work together  to shape the solar   rotation profile in the radiative zone 
 or only one or two are dominant is not settled yet.

Investigations of the impact of rotationally and wave induced transport 
on the structure of stars other than the  Sun have been 
 performed for instance to explain the Li dip  \cite{palacios03}; 
 \cite{charbonnel05}.  
  \cite{Eggenberger05}   studied the  impact of  including both rotationally induced transport
and   magnetic field   on the structure  of   solar like stars  while     
 \cite{Eggenberger10a}  estimate the  seismic consequences and  
 find that with the type of dynamo they assume  in the radiative zone,    
the efficiency of rotational mixing in a radiative zone is significantly decreased
 and seismic parameters  are then similar to those of a non-rotating, non-magnetic star.

\subsection{ Near surface layers}

A direct comparison of the  observed and numerical  frequencies 
of the Sun shows   systematic differences that remain small at low frequency but 
   increase with increasing frequency \cite{jcd88}.  
Several causes contribute to this offset with more or less importance \cite{jcd97}.
 They are collectively referred to  as  
near-surface effects    (for reviews, see for instance \cite{goupil07}, \cite{jcdhoudek10}).

\subsubsection{Surface turbulent  convection}
One important  contribution to the  differences between
the observed and calculated frequencies 
comes from    current modelling of the outer turbulent convective layers of the Sun.
The description of the convective outer layers of the Sun in 1D stellar models remains
quite approximate due to our inability to  represent and  implement in a 1D code  a 
3D multiscale  nonlocal  process such as turbulent dynamics \cite{kupka09a}. 
At the solar surface,  turbulent convection  is inefficient and therefore 
strongly dependent on the free parameters
entering the local, 1D  formulation for convection as well as many other 
assumptions in the formulation. 
A comparison between frequencies computed with 
 two of the available formulations (MLT and CM  \cite{Canuto91}) for instance 
  shows that the frequency differences increase with  frequency and reach  up to $0.15 \%$ at a frequency 
$\nu =4.$ mHz i.e.  a frequency offset of about $ 6-10 \mu$Hz  (\cite{basu08}).  \\
Patched models, that is  1D stellar models where the outer layers have been  replaced in a proper way by those 
obtained with 3D numerical simulations,   lead to   frequencies in much better  agreement 
with the observations \cite{Rosenthal99}, \cite{Li02}, \cite{Samadi10}, \cite{Kupka09b}. For reviews, see
\cite{Samadi09}, \cite{Houdek10a}. 
This approach is valuable for studies of individual stars provided great care is taken 
in the patching procedure, but it 
 cannot be used in  a systematic investigation of a large number of stars. Indeed 3D numerical simulations with the required
 quality are quite numerical time consuming and therefore not available in the whole range of effective temperature,
 gravity and chemical composition.

\begin{figure}[t]
\centering
\resizebox*{0.3\hsize}{!}{\includegraphics*{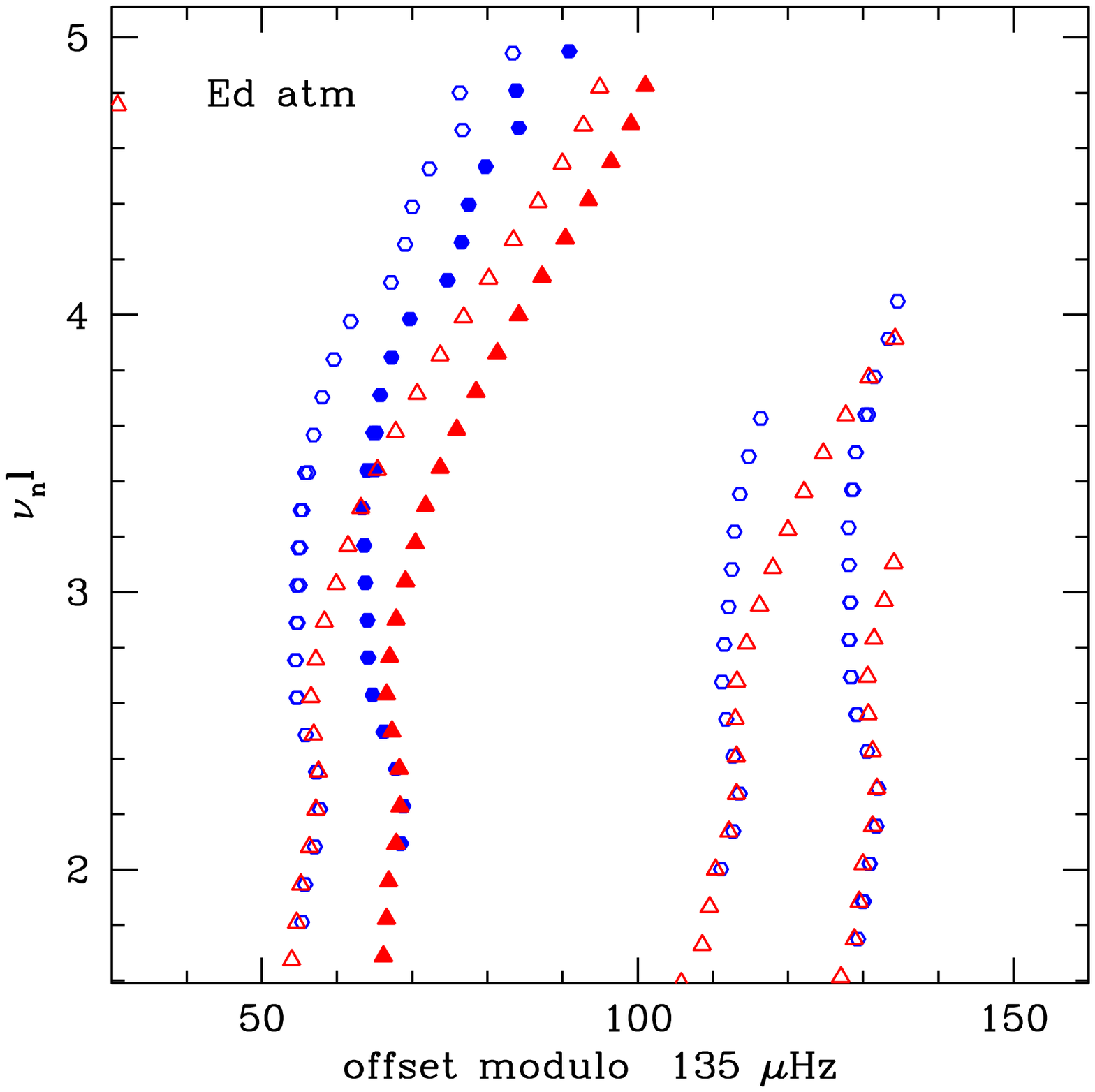}}
\resizebox*{0.3\hsize}{!}{\includegraphics*{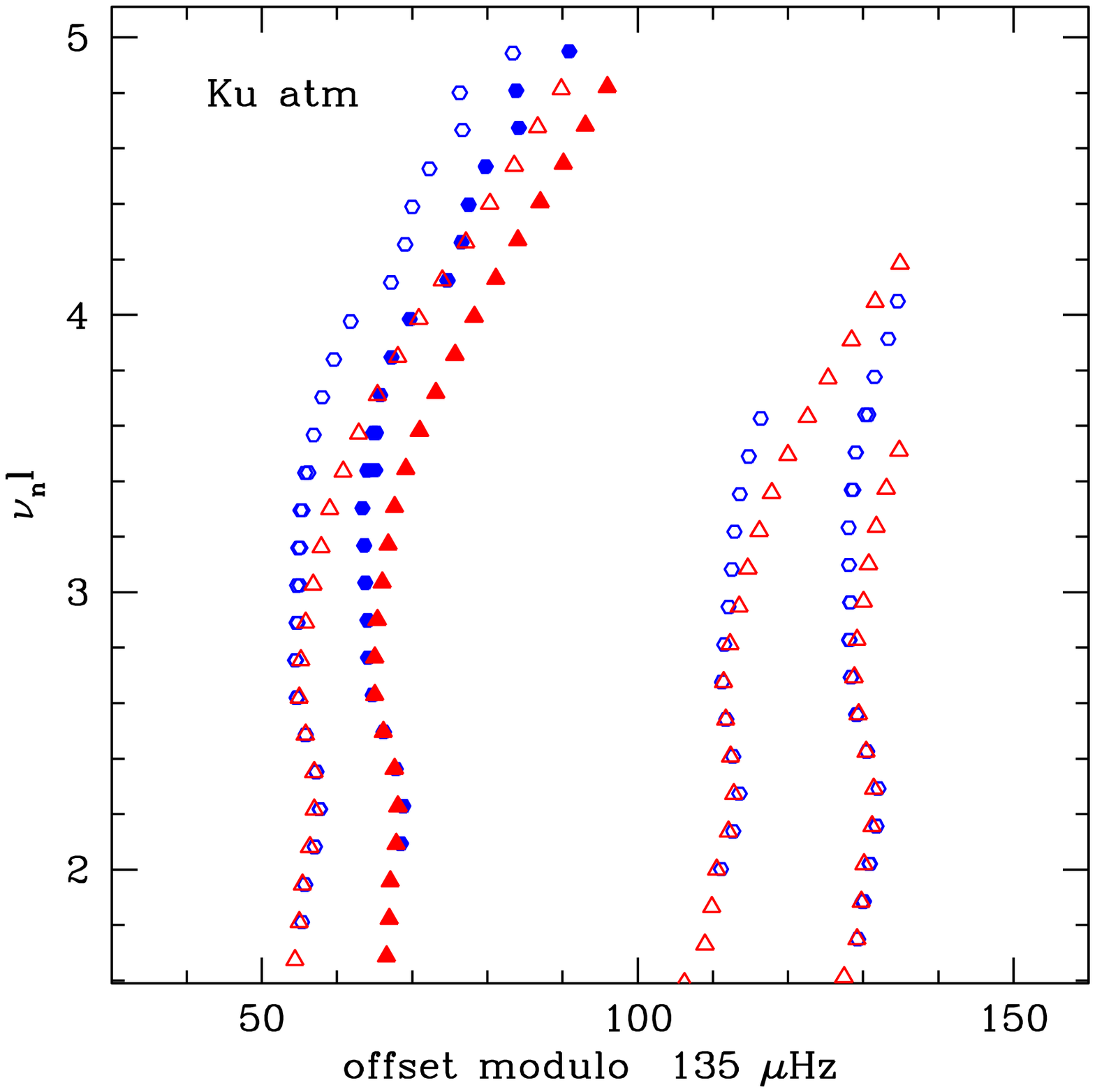}}
\resizebox*{0.3\hsize}{!}{\includegraphics*{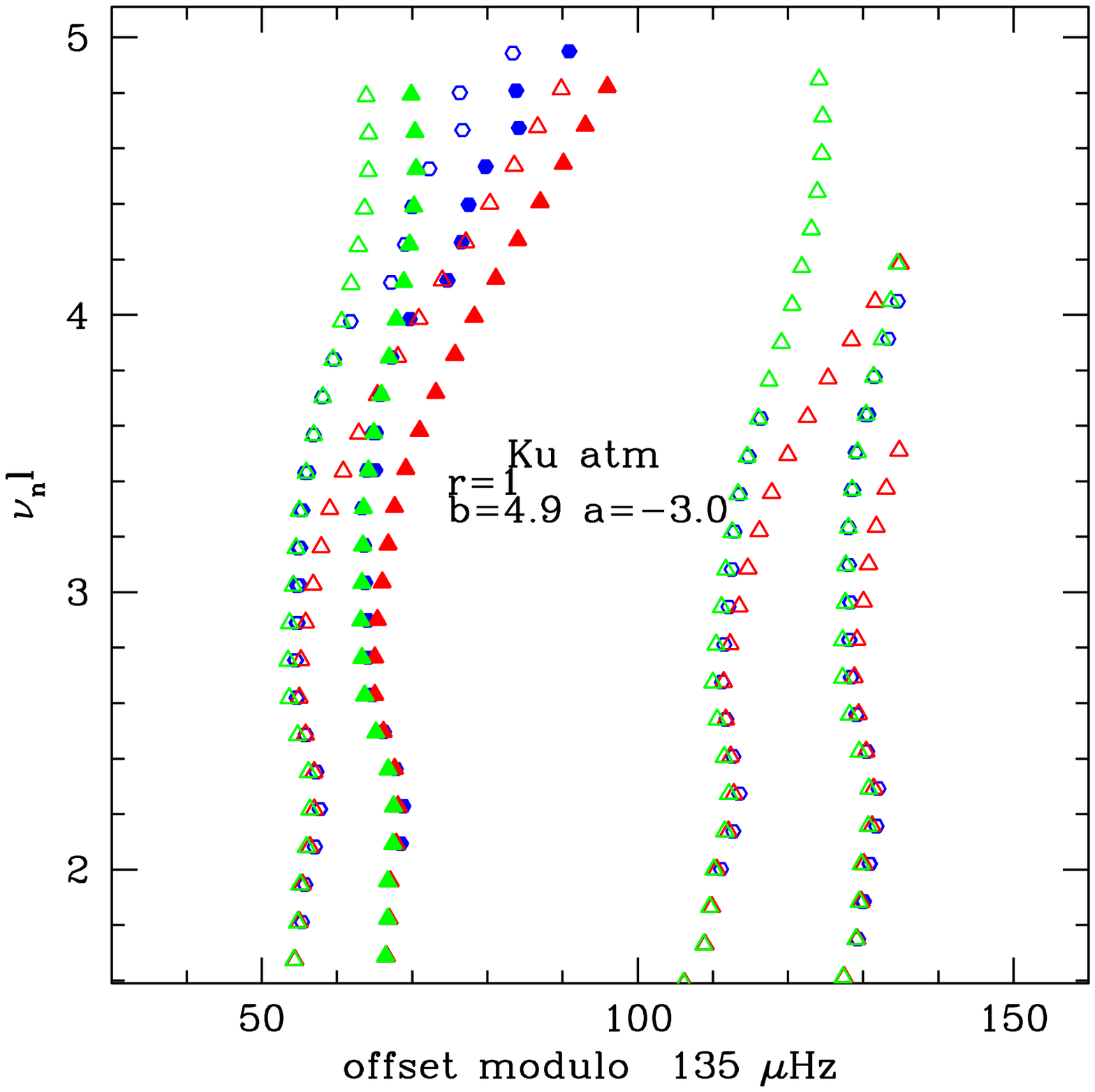}}
\caption{\small  
 Echelle diagrams computed with GOLF solar data (blue dots) and 
   using a calibrated solar model computed with  either an Eddington ({\bf left panel}) or a
    Kurucz  ({\bf middle panel}) $T(\tau)$ law (red dots) . Filled symbols correspond to
radial modes. {\bf Right panel: } An additional  echelle diagram obtained from frequencies of the solar model (red dots) but
  corrected for near surface effects (green dots) agrees with the observations (blue dots) over
   the fitted interval.   
}
\label{echelsun}       
\end{figure}

\paragraph{Atmosphere as boundary condition:}
  Another longstanding  problem is the
description of the atmosphere as boundary condition  when computing a stellar model 
\cite{morel94}. In
the solar case, one can use an empirical atmosphere derived from observations  (HSRA) \cite{gingerich71}
 which represents accurately enough the Sun atmospheric properties. 
 But in other stars, one must rely on 
atmospheres built using a temperature-optical depth, $T(\tau)$, law. Commonly 
 used $T(\tau)$ laws are  the Eddington law  or  more realistic Kurucz model atmospheres \cite{kurucz05a}. 
In order to illustrate the impact of the atmosphere boundary condition, 
Fig.\ref{echelsun} (left panel)   shows an 
echelle diagram  built from observed solar oscillation frequencies  based on  SOHO data. 
An echelle diagram computed for a 
calibrated solar model where the temperature stratification of the 
atmosphere is assumed to follow a Eddington $T(\tau)$ law is also shown (middle panel).
 Largest  differences are seen at high frequency. They are significantly  larger than the observational errors.
The discrepancy at high frequency   decreases (roughly by half) 
when a more realistic Kurucz model is used (middle panel) . 
However  one needs  extensive grids of such models
 along an evolutionary track and for different masses. Furthermore,  model atmospheres 
 suffer from physical imperfections and are not applicable over the entire range of needed
  masses  and ages (i.e. gravity, effective temperature) \cite{kurucz05b}.

\paragraph{Nonadiabatic effects: } Another source of uncertainty comes from what is usually referred to as nonadiabatic
effects. They include the effects of  interaction   of the wave  with
the radiation and with the convection  (see for instance \cite{goupil07}).

\medskip
All these imperfections concur to generate significant errors on the computed oscillation 
frequencies which in the  solar case  amount   up to 10 $\mu$Hz at high frequency. 
As near surface effects cannot yet  be reliably included in theoretical frequency computations,
an alternative has been proposed which consists in removing these effects from  observed frequencies.
In the solar case, this has been
quantitatively assessed  with the comparison of theoretical and observed frequencies of
modes with  small to large  degrees.

 \subsubsection{Correcting frequencies for near surface effects}

In the solar case, due to the large number of different modes, 
near surface effects can be removed from the frequencies.
However,  other procedures must be found for stars other than the Sun. 
Several approaches have been proposed. One consists in comparing the
ratio of frequency combinations  rather than frequencies themselves for low degree modes. In that case 
indeed near surface effects are at least  partially cancelled  \cite{Roxburgh03b}. 
On the other hand,   in order to be able to use  absolute frequencies,
\cite{kjeldsen08} proposed a means to correct observed frequencies  for near surface effects.
They  first showed that the systematic offset between the 
observed and theoretically computed  frequencies of the Sun
 is well   fitted  with a power law
    \begin{equation}
    \label{eqscal}
    \nu_{mod} (n)=\nu_{obs}(n)-a (\nu_{obs}(n)/\nu_{max})^b 
    \end{equation}
with  $a,b$   fit to  the data
and $\nu_{max}= 3100$ Hz is a reference frequency that 
corresponds to the frequency of maximum power in a power spectrum.
The frequencies  $\nu_{obs} (n)$ and $\nu_{mod} (n)$ respectively represent the observed and
  model frequencies of radial modes  with radial order $n$.
For the Sun, such a correction leads for instance to small separations  $d_{02}$ 
 which yield a solar age consistent with the meteoritic age \cite{dogan10}.

Fig.\ref{echelsun} (right panel)  illustrates the effect of correcting $l=0$ 
frequencies from near surface effects according to Eq.\ref{eqscal} with $r=1,a=-3.0,b=4.9  $
for the same solar model  (with Kurucz $T(\tau)$ law) as in the middle panel.
The  corrected frequency echelle diagram coincides with the observations over
   the  frequency interval that was used to derive the parameter values $a,b$.   
The next question is: how much the parameters a,b, $\nu_0$ do depend on the adopted (Kurucz or another) 
model atmosphere?  
Of course one must also keep in mind that  the values of the fitted parameters are 
valid only over the  fitted observed frequency domain (see Fig.\ref{echelsun} left).

\begin{figure}[t]
\centering
\resizebox*{0.45\hsize}{!}{\includegraphics*{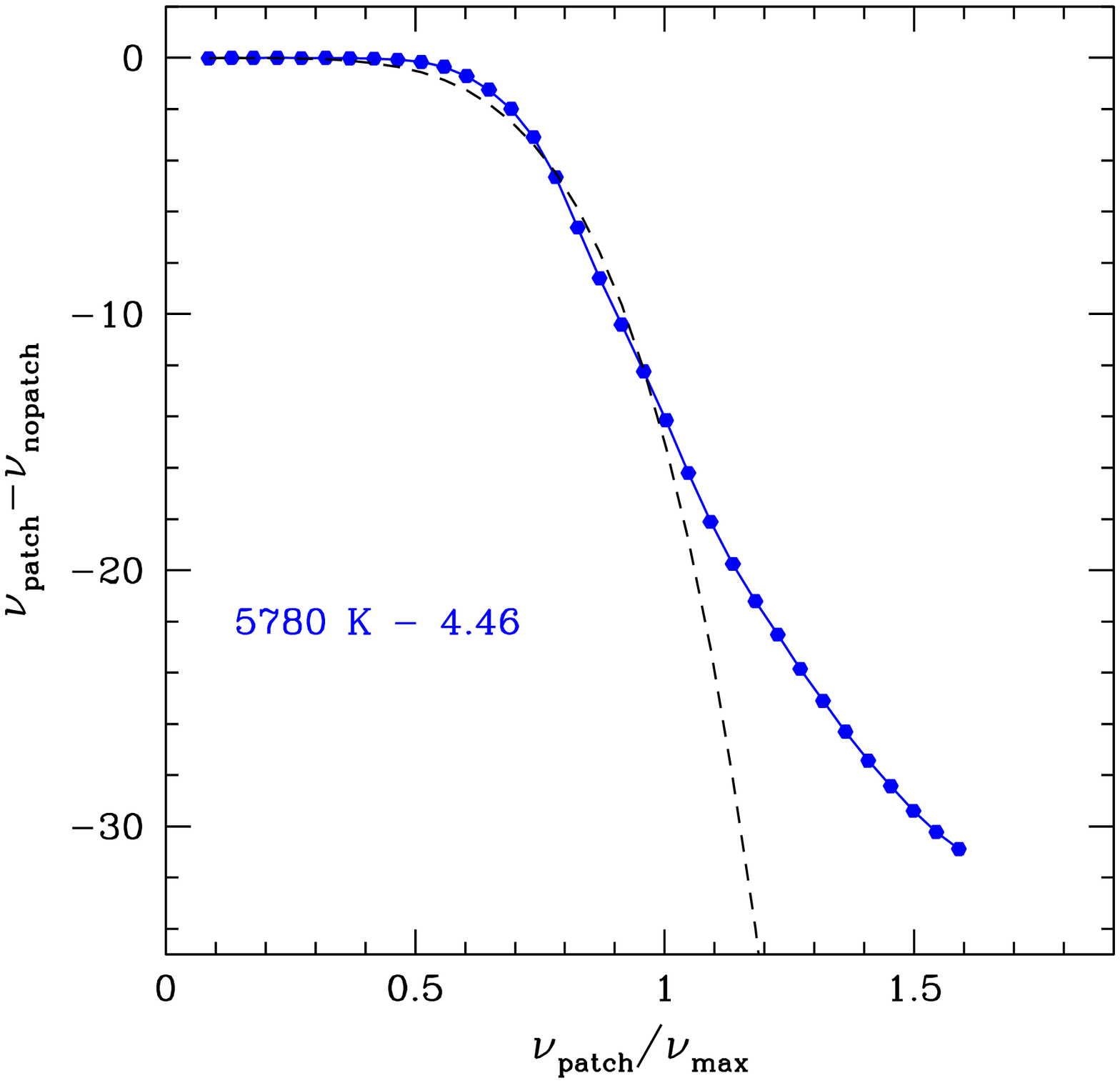}}
\resizebox*{0.45\hsize}{!}{\includegraphics*{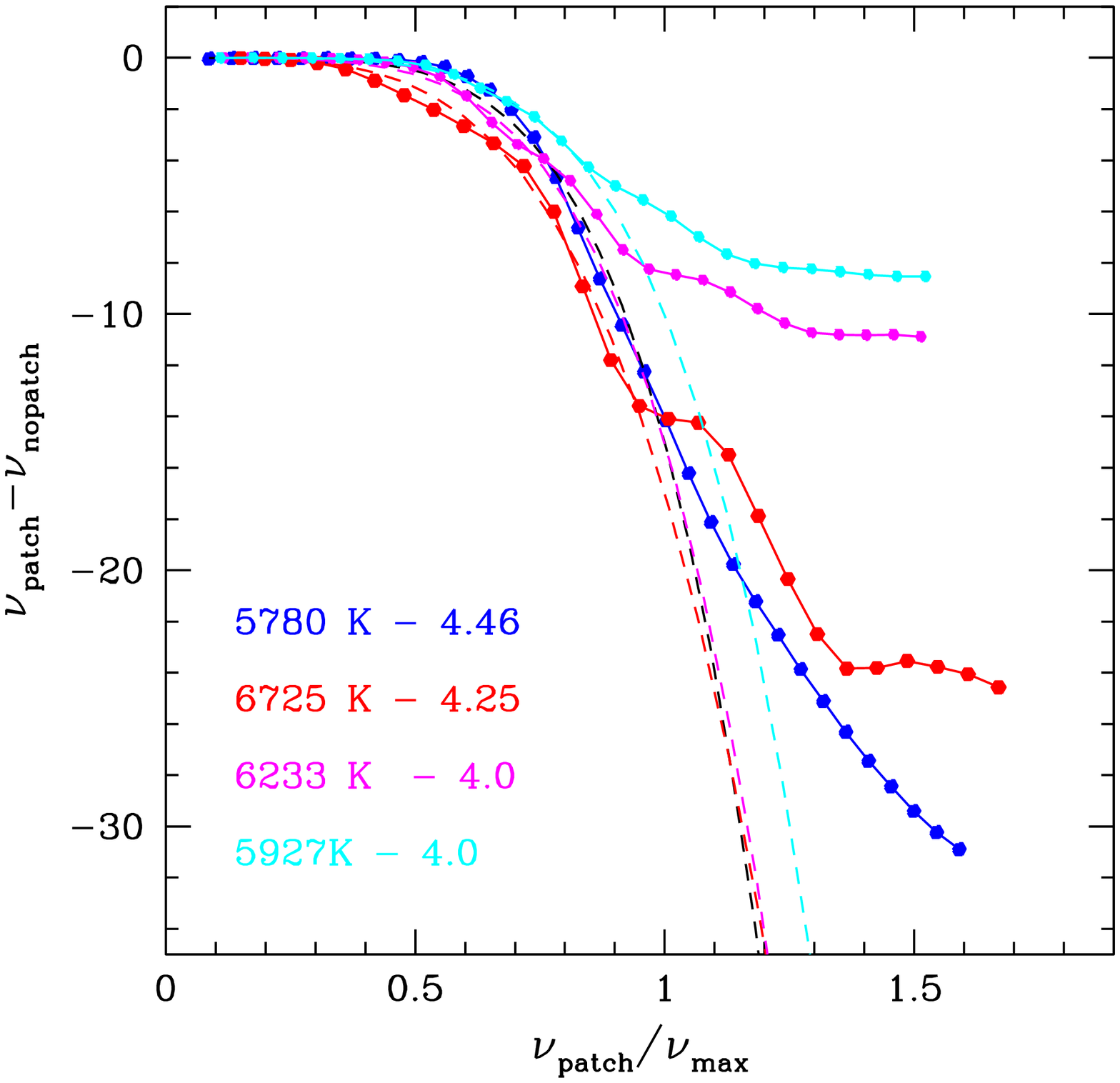}}
\caption{\small {\bf left:}   Frequency differences for radial modes $\nu_{patch}-\nu_{nopatch}$ 
between a patched and a non patched solar models  (blue dots) as a function of $\nu_{patch}/\nu_{max}$.  
The black dashed line represents the power law according to Eq. \ref{eqscal} 
with $\nu_{max}=3.207$mHz,$r=1,a=-15,b=4.9$ (For the solar models: $T_{eff} =5780, g=4.46$). {\bf right:}
Same as left for stellar models with  
different effective temperature, gravity as labelled in the left corner.  
Magenta and red models have a solar metallicity   whereas 
the cyan models have a lower metallicity $[Fe/H]=-1$.   
Coloured dashed lines  represent the power law according to Eq. \ref{eqscal} for each corresponding
frequency difference. Parameter values are $\nu_{max}=1.462$mHz, $r=1,a=-17,b=3.9$ (red); 
$\nu_{max}=1.098$mHz, $r=1,a=-10,b=4.9$ (violet); and $\nu_{max}=1.071$mHz, $r=1,a=-15,b=4.5$ (cyan)  }
\label{diffnu}       
\end{figure}

 Whether the above procedure  can be applied to stars that are different from 
the Sun is an open question. 
Elements of the answer   can be obtained with the use of available 3D numerical simulations 
 and 1D patched models. Frequency differences  for radial modes  between 
a solar patched model and a  non patched model are displayed in Fig.\ref{diffnu} (left) as a function of the 
radial mode frequency of the patched model scaled to the frequency $\nu_{max,\odot}$ given according 
to  \cite{bedding03}  scaling law. The 
  curve  $a(\nu_{patch}/\nu_{max})^b $   is also plotted and coincides
with the observations over   the  frequency interval that was used to derive the  $a,b$ parameter values over an  interval up to $\nu_{patch}/\nu_{max}\approx 1.2$. 
Similar  plots are displayed in Fig.\ref{diffnu} (right) for 3 other models   with  
different effective temperature, gravity or chemical abundance.
For these models, we take again $\nu_{max}$ according to \cite{bedding03} 
and $T_{eff,\odot}$ and $\nu_{max,\odot}$ from  our solar patched model.
 The parameters $a,b$ are  adapted to fit the
frequency differences over the largest  $\nu_{patch}/\nu_{max}$ interval. 
Again a power law can fit the
frequency differences over an interval up to $\nu_{patch}/\nu_{max}\approx 1.2-1.4$.
We note that the frequency differences for the  3 models behave  differently at high frequency
 than the solar ones. This is likely due to their lower gravity, higher effective temperature.
 Indeed,  for these models, the ratio of
turbulent to total pressure is higher than for the solar model (the higher the effective temperature,
 the lower the gravity, the
higher the ratio $P_{turb}/P_{tot} $)   and this results in 
larger differences between patched and non patched models.
   Note that for the three models (all three hotter than the solar model), the
 frequency differences  show an oscillating behavior
  in function of the scaled frequency.  
   Whether this behavior  is
 real or  artificially introduced by the patching process is not known yet. 
In any case,  at lower frequency,  the mean variation with frequency is
 well reproduced by a power law  up to 
 $\nu_{patch}/\nu_{max}\approx 1.2-1.4$

\section{ Conclusion:}

Despite the huge amount of  information provided by helioseismology about the internal structure of the Sun, 
 several important open issues remain that we have reviewed. They  can also impact our understanding and
 modelling  of stars that have a similar structure to that of the Sun.  An increasing set of such stars are 
 being  observed   by CoRoT and Kepler,   and 
  seismic tools developped  for studying  the Sun  are now being adapted to study  other stars.  
    
 \ack { We gratefully thank  our colleagues 
S. Talon,  T. Corbard and  A. Baglin for providing useful information and 
 fruitful discussions when preparing this review. We also thank 
  P. Morel and J. Christensen-Dalsgaard for providing public
 evolutionary and oscillation codes respectively  that were used in the present review. We acknowledge financial
 support from CNES and the  ANR SIROCO}

\section*{References}
\bibliographystyle{iopart-num}
\bibliography{aix_v12a}

\providecommand{\newblock}{}
\begin{thebibliography}{10}
\expandafter\ifx\csname url\endcsname\relax
  \def\url#1{{\tt #1}}\fi
\expandafter\ifx\csname urlprefix\endcsname\relax\def\urlprefix{URL }\fi
\providecommand{\eprint}[2][]{\url{#2}}

\bibitem{baglin06}
{Baglin} A, {Auvergne} M, {Barge} P, {Deleuil} M, {Catala} C, {Michel} E,
  {Weiss} W and {The COROT Team} 2006 {\em ESA Special Publication\/} ({\em ESA
  Special Publication\/} vol 1306) ed {M~Fridlund, A~Baglin, J~Lochard, \&
  L~Conroy} p~33

\bibitem{borucki07}
{Borucki} W~J, {Koch} D~G, {Lissauer} J and {et al} 2007 {\em Transiting
  Extrapolar Planets Workshop\/} ({\em Astronomical Society of the Pacific
  Conference Series\/} vol 366) ed {C~Afonso, D~Weldrake, \& T~Henning} p 309

\bibitem{jcd09}
{Christensen-Dalsgaard} J 2009 {\em {HELAS Workshop on New insights into the
  Sun}\/} ed {M S Cunha and M J Thompson} (\textit{Preprint}
  \eprint{0912.1405})

\bibitem{jcdhoudek10}
{Christensen-Dalsgaard} J and {Houdek} G 2010 {\em \apss\/} {\bf 328} 51--66

\bibitem{basuantia08}
{Basu} S and {Antia} H~M 2008 {\em \physrep\/} {\bf 457} 217--283

\bibitem{jcd02}
{Christensen-Dalsgaard} J 2002 {\em RvMP\/} {\bf 74} 1073--1129

\bibitem{basu2000}
{Basu} S, {Pinsonneault} M~H and {Bahcall} J~N 2000 {\em \apj\/} {\bf 529}
  1084--1100

\bibitem{grevesse93}
{Grevesse} N and {Noels} A 1993 {\em Origin and Evolution of the Elements\/} ed
  {N~Prantzos, E~Vangioni-Flam, \& M~Casse} pp 15--25

\bibitem{basu08}
{Basu} S 2008 {\em 14th Cambridge Workshop on Cool Stars, Stellar Systems, and
  the Sun\/} ({\em Astronomical Society of the Pacific Conference Series\/} vol
  384) ed {G~van Belle} p~10

\bibitem{asplund09}
{Asplund} M, {Grevesse} N, {Sauval} A~J and {Scott} P 2009 {\em \araa\/} {\bf
  47} 481--522

\bibitem{caffau08}
{Caffau} E, {Ludwig} H, {Steffen} M, {Ayres} T~R, {Bonifacio} P, {Cayrel} R,
  {Freytag} B and {Plez} B 2008 {\em \aap\/} {\bf 488} 1031--1046

\bibitem{basu97}
{Basu} S, {Christensen-Dalsgaard} J, {Chaplin} W~J and {et al} 1997 {\em
  \mnras\/} {\bf 292} 243

\bibitem{morel08}
{Morel} P and {Lebreton} Y 2008 {\em \apss\/} {\bf 316} 61--73

\bibitem{jcd96}
{Christensen-Dalsgaard} J, {D\"appen} W, {Ajukov} S and { et al} 1996 {\em
  Science\/} {\bf 272} 1286--1292

\bibitem{bahcall05}
{Bahcall} J~N, {Basu} S, {Pinsonneault} M and {Serenelli} A~M 2005 {\em \apj\/}
  {\bf 618} 1049--1056

\bibitem{Badnell05}
{Badnell} N~R, {Bautista} M~A, {Butler} K, {Delahaye} F, {Mendoza} C, {Palmeri}
  P, {Zeippen} C~J and {Seaton} M~J 2005 {\em \mnras\/} {\bf 360} 458--464

\bibitem{basu10d}
{Basu} S 2010 {\em Proc. of GONG 2010 - SoHO 24 conference: A new era of
  seismology of the Sun and solar-like stars\/} ({\em ESA Special
  Publication\/} vol in press) ed Appourchaux T

\bibitem{turck10d}
{Turck-Chi\`eze} S 2010 {\em Proc. of GONG 2010 - SoHO 24 conference: A new era
  of seismology of the Sun and solar-like stars\/} ({\em ESA Special
  Publication\/} vol in press) ed Appourchaux T

\bibitem{guzik06}
{Guzik} J~A 2006 {\em {Proceedings of SOHO 18/GONG 2006/HELAS I, Beyond the
  spherical Sun}\/} ({\em ESA Special Publication\/} vol 624)

\bibitem{guzik08}
{Guzik} J~A 2008 {\em \memsai\/} {\bf 79} 481

\bibitem{basu07}
{Basu} S, {Chaplin} W~J, {Elsworth} Y, {New} R, {Serenelli} A~M and {Verner}
  G~A 2007 {\em \apj\/} {\bf 655} 660--671

\bibitem{fabbian10}
{Fabbian} D, {Khomenko} E, {Moreno-Insertis} F and {Nordlund} {\AA} 2010 {\em
  ArXiv e-prints\/} (\textit{Preprint} \eprint{1006.0231})

\bibitem{weiss08}
{Weiss} A 2008 {\em PhST\/} {\bf 133} 4025

\bibitem{adelberger10}
{Adelberger} E~G, {Balantekin} A~B, {Bemmerer} D and {et al} 2010 {\em ArXiv
  e-prints\/} (\textit{Preprint} \eprint{1004.2318})

\bibitem{lebreton10b}
{Lebreton} Y 2010 {\em ELSA conference 2010, Gaia: at the frontiers of
  astrometry, S\`evres\/} ed {C Turon, F Arenou, F Meynadier, EAS Series, EDP
  Sciences, in press}

\bibitem{costantini09}
{Costantini} H, {Formicola} A, {Imbriani} G, {Junker} M, {Rolfs} C and
  {Strieder} F 2009 {\em RPPh\/} {\bf 72} 086301

\bibitem{lebreton10a}
{Lebreton} Y and {Montalb{\'a}n} J 2010 {\em \apss\/} {\bf 328} 29--38

\bibitem{DeglInnocenti98}
{degl'Innocenti} S, {Fiorentini} G and {Ricci} B 1998 {\em PhLB\/} {\bf 416}
  365--368

\bibitem{weiss01}
{Weiss} A, {Flaskamp} M and {Tsytovich} V~N 2001 {\em \aap\/} {\bf 371}
  1123--1127

\bibitem{Salpeter54}
{Salpeter} E~E 1954 {\em AuJPh\/} {\bf 7} 373

\bibitem{shaviv2000}
{Shaviv} G and {Shaviv} N~J 2000 {\em \apj\/} {\bf 529} 1054--1069

\bibitem{shaviv10}
{Shaviv} G 2010 {\em \memsai\/} {\bf 81} 77

\bibitem{mussack10}
{Mussack} K and {D{\"a}ppen} W 2010 {\em \apss\/} {\bf 328} 153--156

\bibitem{mao09}
{Mao} D, {Mussack} K and {D\"appen} W 2009 {\em \apj\/} {\bf 701} 1204--1208

\bibitem{maeder09a}
{Maeder} A 2009 vol ~ISBN 978-3-540-76948-4 ({~Springer Berlin Heidelberg})

\bibitem{palacios06}
{Palacios} A, {Talon} S, {Turck-Chi{\`e}ze} S and {Charbonnel} C 2006 {\em
  Proceedings of SOHO 18/GONG 2006/HELAS I, Beyond the spherical Sun\/} ({\em
  ESA Special Publication\/} vol 624) p~38

\bibitem{yang06}
{Yang} W~M and {Bi} S~L 2006 {\em \aap\/} {\bf 449} 1161--1168

\bibitem{turck10}
{Turck-Chi{\`e}ze} S, {Palacios} A, {Marques} J~P and {Nghiem} P~A~P 2010 {\em
  \apj\/} {\bf 715} 1539--1555

\bibitem{Talon04a}
{Talon} S 2004 {\em Stellar Rotation\/} ({\em IAU Symposium\/} vol 215) ed
  {A~Maeder \& P~Eenens} p 336

\bibitem{talon05}
{Talon} S and {Charbonnel} C 2005 {\em \aap\/} {\bf 440} 981--994

\bibitem{maeder09b}
{Maeder} A, {Meynet} G, {Georgy} C and {Ekstr{\"o}m} S 2009 ({\em IAU
  Symposium\/} vol 259) pp 311--322

\bibitem{palacios03}
{Palacios} A, {Talon} S, {Charbonnel} C and {Forestini} M 2003 {\em \aap\/}
  {\bf 399} 603--616

\bibitem{charbonnel05}
{Charbonnel} C and {Talon} S 2005 ({\em EAS Publications Series\/} vol~17) ed
  {G~Alecian, O~Richard, \& S~Vauclair} pp 167--176

\bibitem{Eggenberger05}
{Eggenberger} P, {Maeder} A and {Meynet} G 2005 {\em \aap\/} {\bf 440} L9--L12

\bibitem{Eggenberger10a}
{Eggenberger} P, {Meynet} G, {Maeder} A, {Miglio} A, {Montalban} J, {Carrier}
  F, {Mathis} S, {Charbonnel} C and {Talon} S 2010 {\em \aap\/} {\bf 519} A116

\bibitem{jcd88}
{Christensen-Dalsgaard} J, {Dappen} W and {Lebreton} Y 1988 {\em \nat\/} {\bf
  336} 634--638

\bibitem{jcd97}
{Christensen-Dalsgaard} J and {Thompson} M~J 1997 {\em \mnras\/} {\bf 284}
  527--540

\bibitem{goupil07}
{Goupil} M~J and {Dupret} M~A 2007 ({\em EAS Publications Series\/} vol~26) ed
  {C~W~Straka, Y~Lebreton, \& M~J~P~F~G~Monteiro} pp 93--110

\bibitem{kupka09a}
{Kupka} F 2009 {\em Interdisciplinary Aspects of Turbulence\/} ({\em Lecture
  Notes in Physics, Berlin Springer Verlag\/} vol 756) ed {W~Hillebrandt \&
  F~Kupka} p~49

\bibitem{Canuto91}
{Canuto} V~M and {Mazzitelli} I 1991 {\em \apj\/} {\bf 370} 295--311

\bibitem{Rosenthal99}
{Rosenthal} C~S, {Christensen-Dalsgaard} J, {Nordlund} {\AA}, {Stein} R~F and
  {Trampedach} R 1999 {\em \aap\/} {\bf 351} 689--700

\bibitem{Li02}
{Li} L~H, {Robinson} F~J, {Demarque} P, {Sofia} S and {Guenther} D~B 2002 {\em
  \apj\/} {\bf 567} 1192--1201

\bibitem{Samadi10}
{Samadi} R, {Ludwig} H, {Belkacem} K, {Goupil} M~J and {Dupret} M 2010 {\em
  \aap\/} {\bf 509} A15

\bibitem{Kupka09b}
{Kupka} F, {Belkacem} K, {Goupil} J and {Samadi} R 2009 {\em Co.Ast.\/} {\bf
  159} 24--26

\bibitem{Samadi09}
{Samadi} R 2009 {\em ArXiv e-prints\/}

\bibitem{Houdek10a}
{Houdek} G 2010 {\em \apss\/} {\bf 328} 237--244

\bibitem{morel94}
{Morel} P, {van't Veer} C, {Provost} J, {Berthomieu} G, {Castelli} F, {Cayrel}
  R, {Goupil} M~J and {Lebreton} Y 1994 {\em \aap\/} {\bf 286} 91--102

\bibitem{gingerich71}
{Gingerich} O, {Noyes} R~W, {Kalkofen} W and {Cuny} Y 1971 {\em \solphys\/}
  {\bf 18} 347--365

\bibitem{kurucz05a}
{Kurucz} R~L 2005 {\em \memsai\/} {\bf 8} 14

\bibitem{kurucz05b}
{Kurucz} R~L 2005 {\em \memsai\/} {\bf 8} 73

\bibitem{Roxburgh03b}
{Roxburgh} I~W and {Vorontsov} S~V 2003 {\em \aap\/} {\bf 411} 215--220

\bibitem{kjeldsen08}
{Kjeldsen} H, {Bedding} T~R and {Christensen-Dalsgaard} J 2008 {\em \apjl\/}
  {\bf 683} L175--L178

\bibitem{dogan10}
{Do{\u g}an} G, {Bonanno} A and {Christensen-Dalsgaard} J 2010 {\em ArXiv
  e-prints\/} (\textit{Preprint} \eprint{1004.2215})

\bibitem{bedding03}
{Bedding} T~R and {Kjeldsen} H 2003 {\em \pasa\/} {\bf 20} 203--212

\end{thebibliography}

\end{document}